\documentclass[5p,times,twocolumn]{elsarticle}

\usepackage{lineno,hyperref}
\modulolinenumbers[5]

\journal{a journal}

\begin{document}

\begin{frontmatter}
\title{A new concept of a high-energy space-based cosmic ray telescope}

\author{Carmelo Sgr\`o}
\ead{carmelo.sgro@pi.infn.it}
\author{Marco Incagli}
\address{INFN-Pisa, Largo B. Pontecorvo 3, 56127 Pisa, Italy}

\begin{abstract}
  Cosmic ray science has proved to be a very active field, with several
 important results from recent space-based instruments.
  Next generation experiments will explore the multi-TeV energy range,
  trying to cope with the mass and power budget constraints of artificial
  satellites that limits the collecting area thus reducing the available
  statistics at the highest energy.
  With the aim to improve the compromise between area and mass, we propose
  a new concept for a cosmic-ray telescope in which the detector elements
  are organized in bars along the 3 axis. In this way we can also maintain
  a good event shower sampling (for direction and energy reconstruction)
  and a relatively small number of channels (required power)
  as the detector size increases. A possible implementation
  of the concept is also evaluated with a Geant4 simulation.
\end{abstract}


\end{frontmatter}

%%%%\linenumbers

\section{Introduction}

The high-energy cosmic ray field has grown significantly in recent years.
This generation of space-based detectors, like 
Pamela~\cite{Pamela}, AMS-02~\cite{AMS02} and Fermi-LAT~\cite{LATPaper},
provided new, unexpected results that, while increasing significantly 
our knowledge of characteristics of primary cosmic rays, 
posed also new questions. 
As an example, diffusive propagation models, that work nicely for most 
of the observed spectrum, are considered  less reliable above few TeV
where we expect to see effects of e.g. local sources.
In particular, high-energy cosmic-ray electrons and positrons
radiate energy very quickly 
and carry information only of the nearby part of our Galaxy. 
Their spectrum and, possibly, their anisotropy in incoming direction,
can lead to the discovery of acceleration sources, but high statistic 
measurement at energy above 1 TeV are required. 
Hints of dark matter are expected to be seen in $\gamma$-rays and cosmic rays,
but current uncertainty on the astrophysical component 
(that can be considered as background in such searches) is one 
of the main limitation.
New measurements in an extended energy range with larger statistics
and smaller uncertainty are thus mandatory.

The next generation of detectors are trying to answer all these question
by improving the measurement capability in the few TeV range, 
where the performance of the current detectors are limited. 
Recently launched instrument like CALET~\cite{CALET} or DAMPE~\cite{DAMPE} 
chose to have a very deep calorimeter.
The good energy resolution is achieved at the price of a smaller acceptance:
the geometry factor is of the order of a 0.2--0.3 m$^2$ sr for electrons, 
which has impact on the time needed to collect enough statistics above 1 TeV.
On the contrary the Fermi-LAT, optimized for $\gamma$-ray in 
the 1--100 GeV range, made the opposite choice with a large geometry factor 
of the order of 2 m$^2$ sr, but with a worse energy resolution at 1 TeV.

Magnetic spectrometers naturally suffer of a lower acceptance when compared 
to pure calorimetric experiments of the same mass.
The magnetic system represents heavy passive material, and 
reduces the field of view. On the other hand such device is required 
to measure particle change, in particular to separate positive to negative 
charge. If a good energy resolution for hadron is required, magnetic 
spectrometer can use particle curvature for momentum measurement, but this 
technique works best at lower energy and optimization in the multi TeV is very 
demanding in terms of mass, power and size.

A calorimeter concept has been recently proposed to have an
acceptance of $\sim 3$ m$^2$ sr to collect enough statistics in a reasonable 
time to explore the multi-TeV region~\cite{Calocube}. 
The idea is to build a cubic calorimeter 
made of small cubic sensitive elements in order to have a 3D, deep, 
homogeneous and isotropic calorimeter. The large acceptance is achieved
by accepting events from 5 out of 6 surfaces, while the last one is used
for mechanical and electrical interfaces. 
The cubic element is usually a scintillating crystal,
with the size of the order of the Moli\`ere radius for good shower sampling.
The R\&D on this kind of detector is well advanced.

While this idea is certainly interesting, this geometry requires some
gaps between the modules for the routing of the readout cables and 
the mechanical structure.
Moreover the number of channels is proportional to the volume of the detector,
and scaling up its size means scaling also power consumption proportional 
to the volume.
It is easier (cheaper) to put heavy satellites 
in orbit close to the Earth (Low Earth Orbits or LEO),
therefore the solid Earth and its atmosphere
can partially shield the detector field of view, as a result the gain of
using 5 surfaces instead of 1 is partially reduced.
As a reference the largest launch system currently available can put
about 20 metric tons in LEO, but a few tons is currently considered
a feasible size. As an example AMS-02 is about 7 tons and is probably
the heaviest instrument currently operational.

In this work, we got inspired from the idea of the cubic and modular instrument,
but modified a bit the geometry to try to limit the issues just described. 
The new concept will be fully described in section~\ref{sec:concept}, 
while in the following one we show some calculation of the achievable
acceptance and field of view, in particular in LEO. 
Finally in section~\ref{sec:g4sim} we will describe one possible 
implementation of the detector to prove that we can reach reasonable preformance.
Of course this work is not a full study of a complete detector design, 
but a proposal of a geometrical concept to be further refined in order to
become a feasible and high-performance space-based cosmic-ray telescope.

\section{The concept}\label{sec:concept}

The cosmic-ray telescope we are proposing is a pure calorimetric 
instrument, with emphasis on large collection area or, 
in other words, must be easily scalable to large volumes. 
A good energy resolution is of course important, but it is mainly given 
by the depth of the detector, which obviously scale with its size. 
However it must be capable to measure the energy of 
many type of cosmic rays: electrons, protons, alpha particles, etc.
Granularity of the calorimeter is considered fundamental. 
It allows shower development imaging, for a better energy reconstruction and 
for particle identification. Particle incoming direction can be also measured,
which is important for anisotropy searches. 
A good granularity usually implies a large number of channels, 
which in turn means large power consumption and event size.
This can be a problem for very large detector if the number of channels
is proportional to the volume.

\begin{figure}[hbt]
\includegraphics[width=\linewidth]{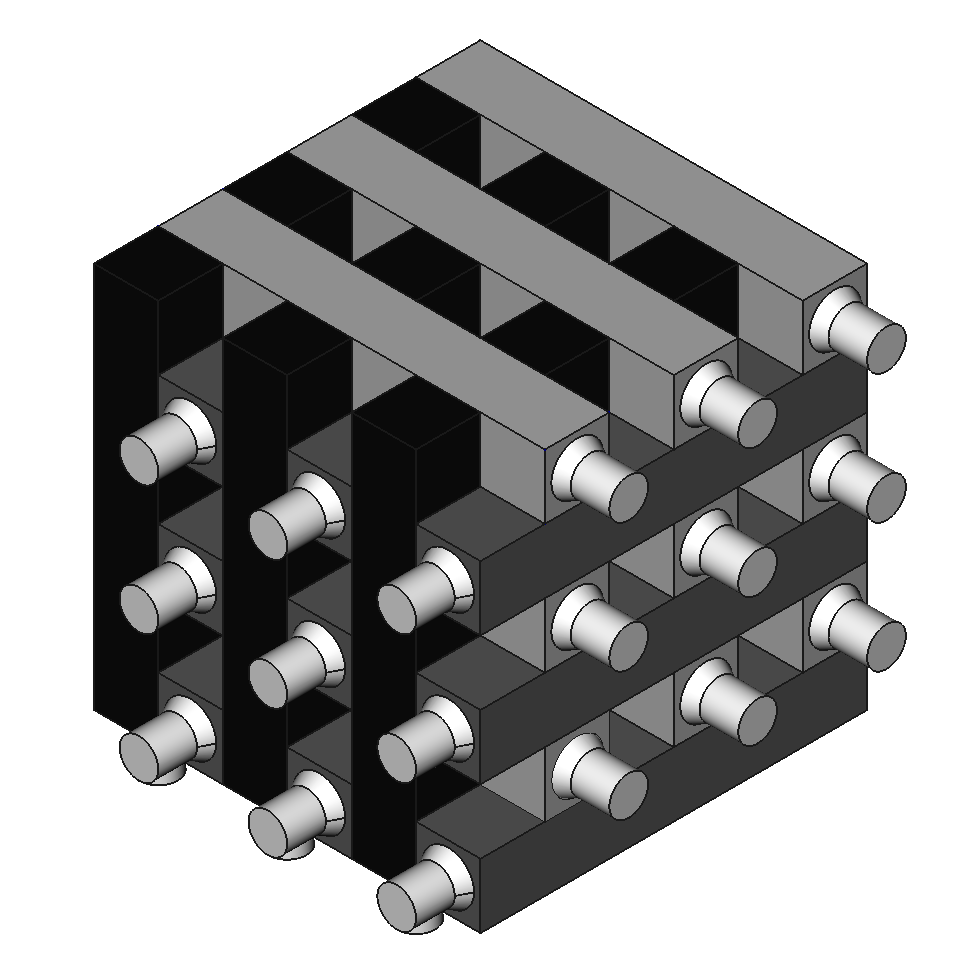}
\caption{Graphical representation of the detector concept, 
here with only 9 bars per side. 
Different colors (gray level) are used for different bar orientations, while 
the white cylinders represent the readout devices.
}\label{fig:CalSchematic}
\end{figure}

The detector in this concept is a cube on side length $L$, with detector
elements that are bars of the same length $L$ and a much smaller side $S$. 
Each of the element is read from one side only, and provides information 
on energy deposition in the bar. The information of the position is given by the
physical place of the bar itself, no information on longitudinal position 
(i.e. along the bar) is strictly required. 
The three dimensional sampling of the shower development inside the detector
is given by arranging three sets of bars along the three coordinates $x,y,z$.
The elements in one set of bars are arranged parallel one another 
in a $n\times n$ grid, and spaced with a pitch that is twice their side;
in this way there is enough empty space to combine the 3 sets 
in a single cubic shape.
Figure~\ref{fig:CalSchematic} shows an example of this concept, using only 
9 elements per side, while in a real detector the number of elements can 
increase arbitrary and the only constraint is that $L = 2\cdot S\cdot n$.
In figure~\ref{fig:CalSchematic} the readout system is shown with 
white cylinders, just to make clear that only 3 out of 6 sides are used
to readout the full instrument. 
The other three, that we will call ``active'', can be instrumented 
with other subsystems for precision tracking, charge Z measurements etc.
To keep the active surfaces on the top of the full assembly the cube can be
rotated to sits on one the corner. Of course the mechanical support structure 
must be designed to use the 3 non active surfaces and the rotated geometry,
but the exact design of the mechanics if out of the scope of this work.

No specific assumption on the detector element technology is done at this point.
It can be either a homogeneous absorption system like an inorganic
scintillator, or a sampling device with two or more materials. 
In section~\ref{sec:g4sim} we will abandon some generality and describe 
one specific case, as an example of the achievable performance. 
In the rest of this section we just want to highlight some of the pros and cons
of this concept, in particular comparing it with the case of the ``regular'' 
cube that sits on one surface, and have the other 5 active.

As mentioned, scaling to large volume is a major drivers, and this design
guaranties that while the volume increase with $L^3$, the 
number of channels is proportional only to $L^2$. 
This is a clear advantage with respect to a design in which 
the number of channels is proportional to the volume 
(i.e. increase faster than the acceptance when $L$ increases).
All the readout devices are on external surfaces (the 3 non active ones, 
possibly at the bottom and close to the spacecraft structure), simplifying 
the routing of the electrical connections. Moreover the detector pitch 
of $2S$ leaves enough space for readout sensors and electronics.

In this design there is some empty space inside the calorimeter:
1/4 of the total volume is not instrumented.
However this is not required for a space-based cosmic-ray telescope and,
it allows for a larger volume (acceptance) if compared to a compact design
with the same mass, since the average density is 3/4 of the detector material.

Particle showers are sampled in the 3 coordinates and it is possible to
reconstruct shower direction and shape. In section~\ref{sec:g4sim} we will see 
how we can reach an angular resolution of few degrees.
We mentioned that shower sampling can be used for leakage correction
and quantify some kind of ``quality'' of the energy measurement. 
However, if the calorimeter is deep enough, it can measure the energy 
via total absorption, and leakage correction with shower shape analysis 
is less important. It remains true that to fully exploit this geometry 
complex reconstruction algorithms can be required, for both direction, energy 
and shower topology.

\section{Detector acceptance}\label{sec:acceptance}
A simple toy Monte Carlo simulation is used to evaluate the effective 
geometry factor, or acceptance, of the concept described in this work. 
Events are extracted from a surface of 25~m$^2$ with isotropic direction 
distribution, and are propagated to detector represented by a cube of edge L.
An event is considered good if it enters from one of the active surfaces and 
if it crosses more than L/2 inside the detector volume. The latter cut 
mimics a minimum quality requirement, and tries to remove events that clips 
the edges of the detector. The exact cut in a real detector can be 
different, but in this simple exercise we don't want to go in such details.

We simulated different cubes of side L from 50 cm to 150 cm, 
for each case we consider both the ``regular'' cube, that sits
on one surface (considered the only non active), and our ``rotated'' geometry
in which the cube sits on a corner and only the top 3 surfaces are active. 
Moreover we consider the case of just the quality cut and with the additional
request of accepting event with local zenith angle $< 100^{\circ}$ to avoid 
events from the Earth Limb in case of satellite in LEO~\footnote{See e.g.
the discussion on Earth Limb in 
\url{http://fermi.gsfc.nasa.gov/ssc/data/analysis/LAT_caveats.html}}. 
As for the quality cut, this one can be different form the actual cut of 
a real instrument (which depends the orbit, the angular resolution, etc), 
but is a reasonable representation of the reduction of field of view from 
the solid Earth.

\begin{figure}[hbt]
\includegraphics[width=\linewidth]{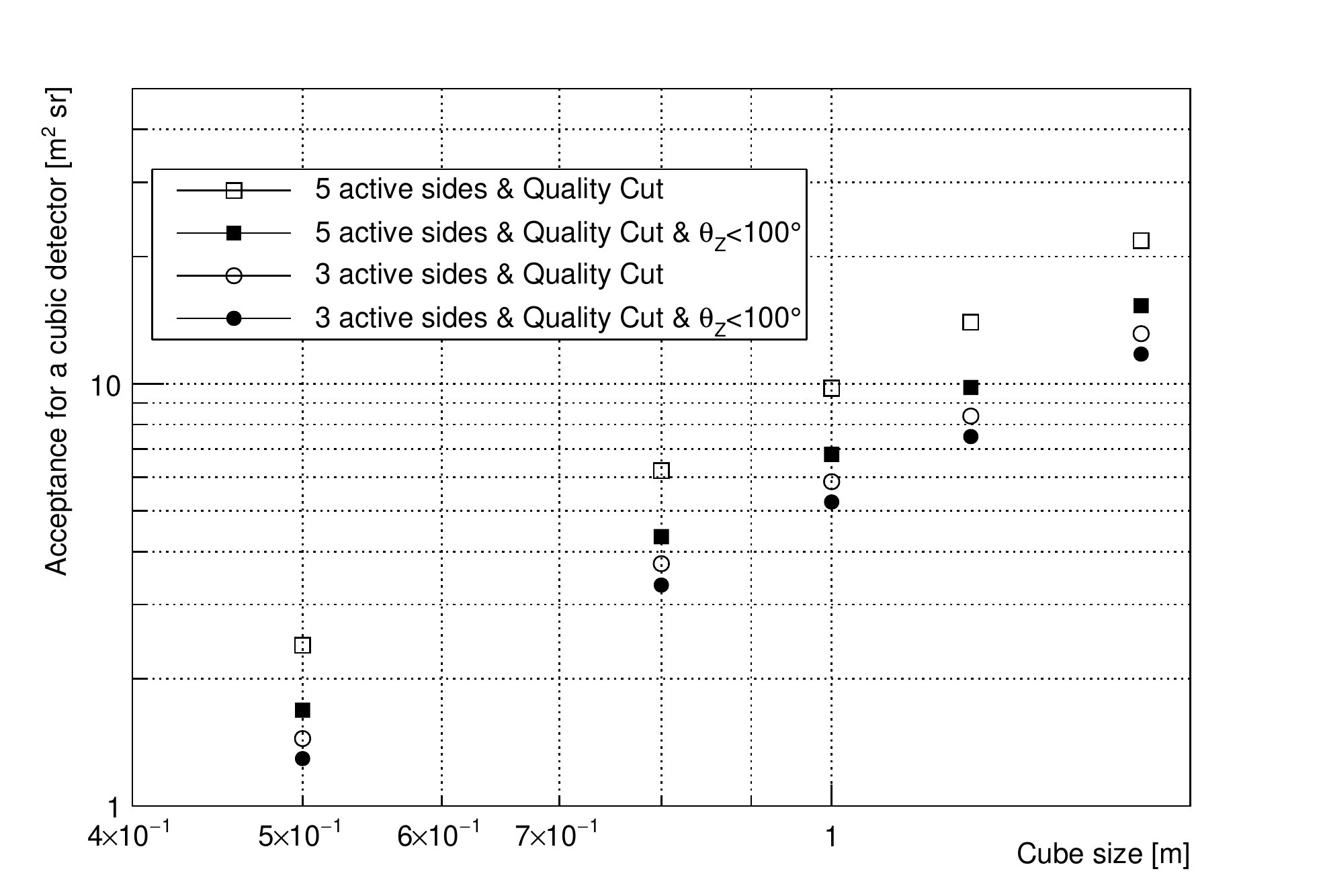}
\caption{Acceptance for cubic detector with 3 and 5 active faces 
as function of cube side. Both case with simple cut and Eart-limited 
field of view are show.
}\label{fig:Aeff}
\end{figure}

The acceptance is defined as the ratio between good 
and total generated events, normalized by area and solid angle. 
The result of the simulation is presented in figure~\ref{fig:Aeff}, 
where we can see that the acceptance increases as the square of the cube size,
and that L $\sim 1$~m is necessary to reach acceptance large enough for 
the multi TeV range. To have an idea on the corresponding mass of such
a detector, we can assume an average density of a commonly used
scintillator like CsI ($4.5~g/cm^3$) which leads to a reasonable mass of
about 4 (8) metric tons at L=1~m (1.2~m) while for the maximum L=1.5~m
considered here we obtain about 15 metric tons; not totally unfeasible,
but pretty close to the limits.

The regular cube, with 5 active surfaces provides 
always superior acceptance than our proposed solution (with only 3 active 
surfaces), but its advantage is only a factor $\sim 1.3$ in a LEO.
In fact the local zenith cut removes about 40\% of the events in the
regular cube case, and about 12\% in our configuration.
These removed events can still be 
considered valuable for calibration\footnote{As an example, events from the
Earth's limb are a source of high energy $\gamma$-rays useful for on orbit 
check of electromagnetic response.} so we think that our geometry provides 
a more reasonable ratio of diagnostic/science datasets.

\begin{figure}[hbt]
\includegraphics[width=\linewidth]{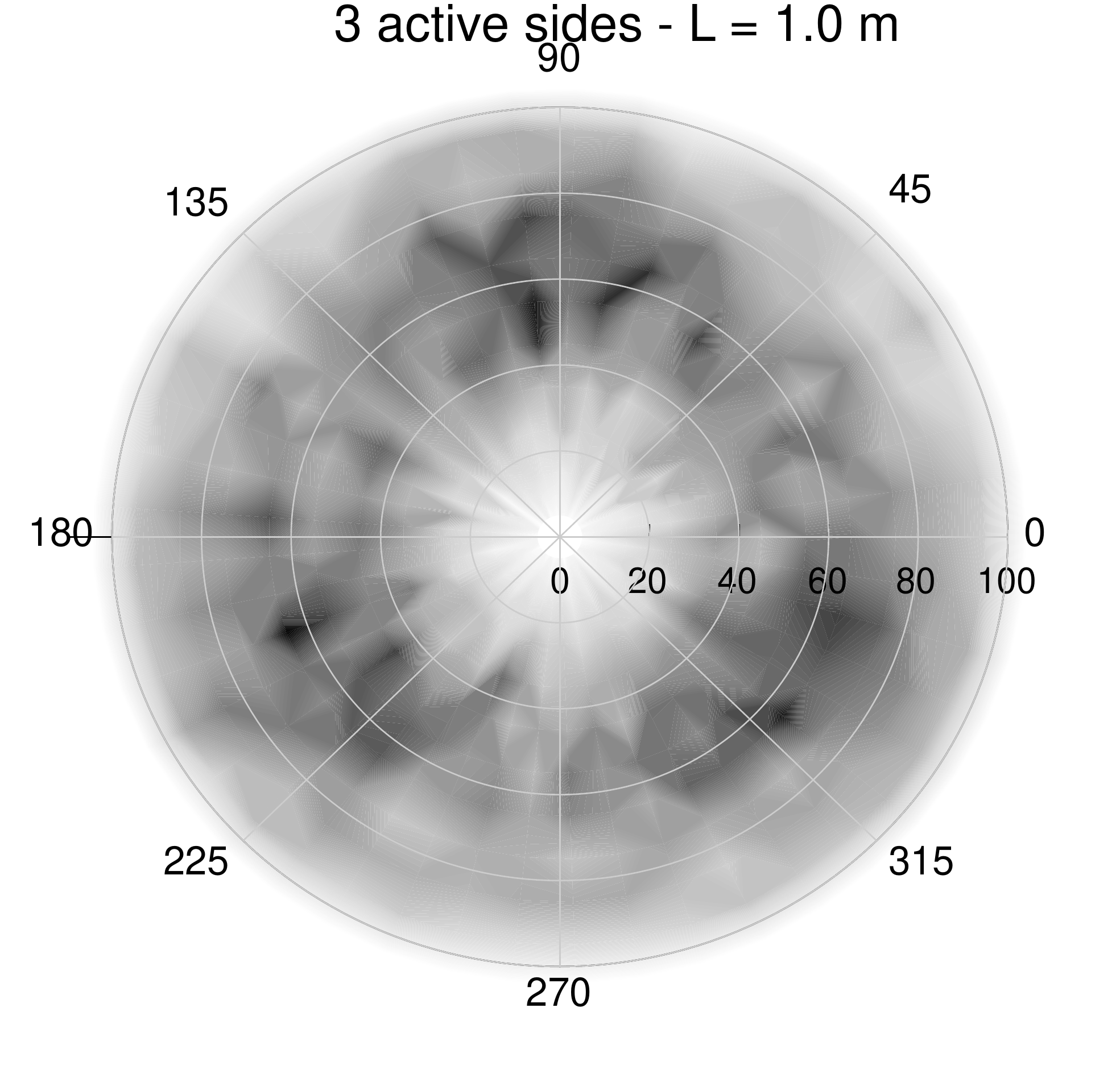}
\caption{Polar plot of the acceptance as function of spherical coordinate 
$\theta-\phi$ in the detector reference frame. The intensity of the color 
is proportional to the acceptance in that direction. 
Here only the case L= 1 m is plotted to show the asymmetry (about 15\%) due to 
the orientation of the 3 active surfaces of the proposed geometry.
}\label{fig:AeffPolar}
\end{figure}

From this simulation, we can also study the acceptance as function of the
polar coordinates $\theta-\phi$ in the detector reference frame, in order 
to study the uniformity of the field of view.
Figure~\ref{fig:AeffPolar} shows an histogram of the number good events 
(thus proportional to the acceptance) as function of $\theta-\phi$.
We can see that the field of view is not completely uniform, 
the 3 active surfaces are clearly visible at $\phi=90^{\circ}$, $210^{\circ}$,
and $330^{\circ}$ in this reference system. 
The asymmetry is about 15\% and does not depend on the cube side L.

\section{Simulation of a simple implementation}\label{sec:g4sim}
The exercise in the previous section involves only a simple cubic geometry,
and does not include any details on the detector elements. 
However to be able to estimate performance like angular and energy resolution 
for both electron and hadrons, we need to take a further step and make a more 
detailed simulation of the detector. This implies making some decisions on 
the detector element technology and detailed geometry (size and length). 
Moreover, some event reconstruction algorithm must be introduced, in order 
to take advantage of the three dimensional sampling of the shower development.

The details of the technology and algorithm can significantly change 
the performance of the instrument. In this work we don't want to propose a 
specific detector layout, nor to optimize the detector
parameters for specific requirements or to implement a full simulation,
including e.g. detailed sensor behavior.
We just want to provide an idea of the expected performance, to be further 
studied and optimized in future works, taking also into account 
available technology and laboratory tests.
We implemented a simple simulation of a possible detector configuration, 
based on Geant4 toolkit~\cite{Geant4}, and we will show the results of 
a few tests with electron and proton beams, using a very simple shower
reconstruction.

\begin{figure}[hbt]
\centering
\includegraphics[width=0.6\linewidth]{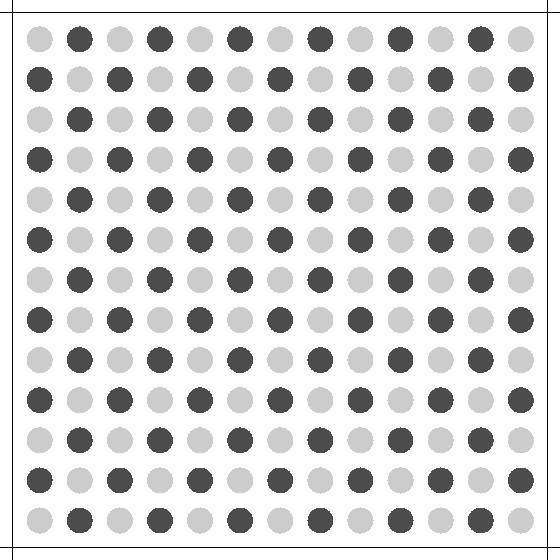}
\caption{Cross section of a calorimeter bars that shows the layout 
of the scintillating (dark gray) and clear fibers (light gray).
}\label{fig:fibers}
\end{figure}

In this exercise we implemented a sampling calorimeter with copper
as absorber material (atomic number 29, density 8.960  g/cm$^3$)
\footnote{
\url{http://pdg.lbl.gov/2015/AtomicNuclearProperties/HTML/copper_Cu.html}}
and scintillating fibers to sample the energy. We also included clear fibers 
to sample the Cherenkov light and to provide a further set 
of information for protons. This choice of material roughly follows
the one done by the RD52~\cite{RD52} proposal for their hadronic calorimeter.
The basic element is a bar 80~cm long and 2 cm wide.
The bars are assembled in a regular grid having $20\times 20$ readout elements
in each ot the 3 faces. See figure~\ref{fig:CalSchematic} for a sketch of
a $3\times 3$ matrix.
In each bar the sampling is done with a grid of
$13\times 13$ fibers of 1 mm diameters, alternating scintillating 
and clear fibers as in figure~\ref{fig:fibers}. 
Scintillating fibers are made of Polystyrene (density 1.05 g/cm$^3$) 
with a 0.01 mm thin layer of PMMA cladding (density 1.19 g/cm$^3$), 
the clear fibers have a core of PMMA (index of refraction 1.49) 
with fluorinated polymer as cladding (index of refraction 1.42).
We chose to limit the bar length to 80~cm to avoid a too computing intensive
simulation, and to keep it closer to the dimensions of a feasible prototype
for laboratory tests.
The mass of the calorimeter (including fibers and empty spaces)
can be estimated in about 2.44 tons, with an average density of 4.77~g/cm$^3$,
not far from CsI one.
By choosing titanium instrad of copper, as passive material,
the average density becomes 2.6~g/cm$^3$,
and even a cube of L=1.5~m would lead to a reasonable mass
of about 9 tons. Table~\ref{tab:material} summarise a few example of possible
material choice.

\begin{table}
\begin{center}
  \begin{tabular}[!h]{l|cc|ccc}
    \hline
    Material &  X$_0$ & L$_{int}$ & $<\rho>$ & M$_{80}$ & M$_{150}$\\
    & [cm] & [cm] & [g/cm$^3$] & [t] & [t] \\
    \hline
    \hline
    Titanium & 3.56 & 27.80 & 2.6  & 1.3 & 8.8\\
    Iron     & 1.76 & 16.8  & 4.2  & 2.2 & 14.2\\
    Copper   & 1.44 & 15.32 & 4.77 & 2.44 & 16.1 \\
    Lead     & 0.56 & 17.6  & 6.0  & 3.0 & 20\\
    \hline
  \end{tabular}
\end{center}

  \caption{Comparison of a few absorber materials: radiation length X$_0$
    and nuclear interaction length L$_{int}$ of each material are shown
    together with the resulting average density $<\rho>$ of the detector.
    For completeness we show also the total mass for a cube of 80~cm
    (M$_{80}$) as in used setup, and of 150~cm (M$_{150}$) the largest
    considered in section~\ref{sec:acceptance}.
    It is worth noticing that commercially available materials are
    usually alloys, but their properties are not far from the main element.
  }\label{tab:material}
\end{table}

The simulation saves the energy in the scintillating fibers in each bar,
the ``S'' signal in the rest of this work, and number of Cherenkov photons 
in clear fibers, the ``C'' signal. 
There is no attempt to simulate a readout system, like electronic 
and/or detector gain and noise. 

\begin{figure*}
\begin{center}
\includegraphics[width=0.8\textwidth]{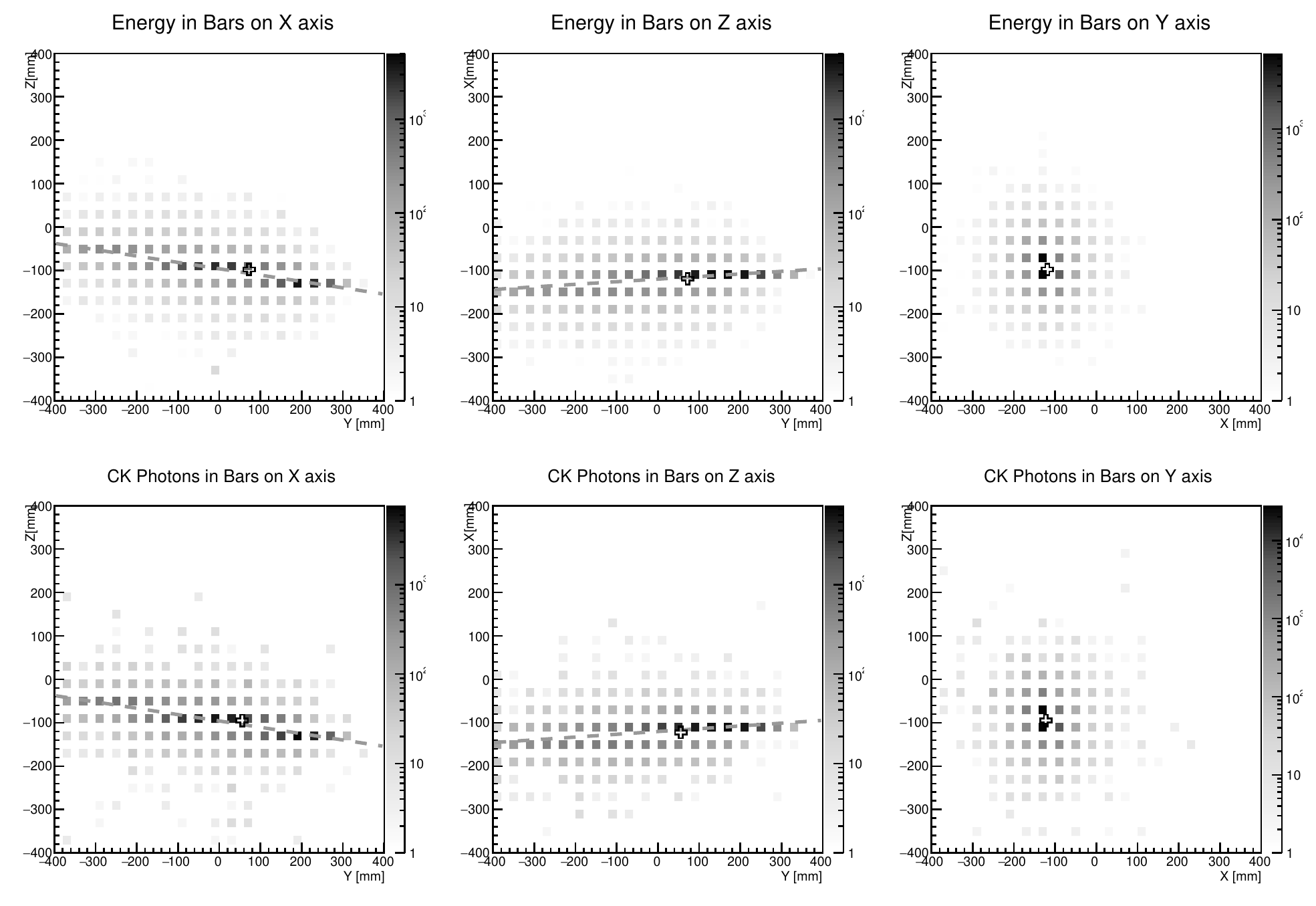}\\
\hrule
\includegraphics[width=0.8\textwidth]{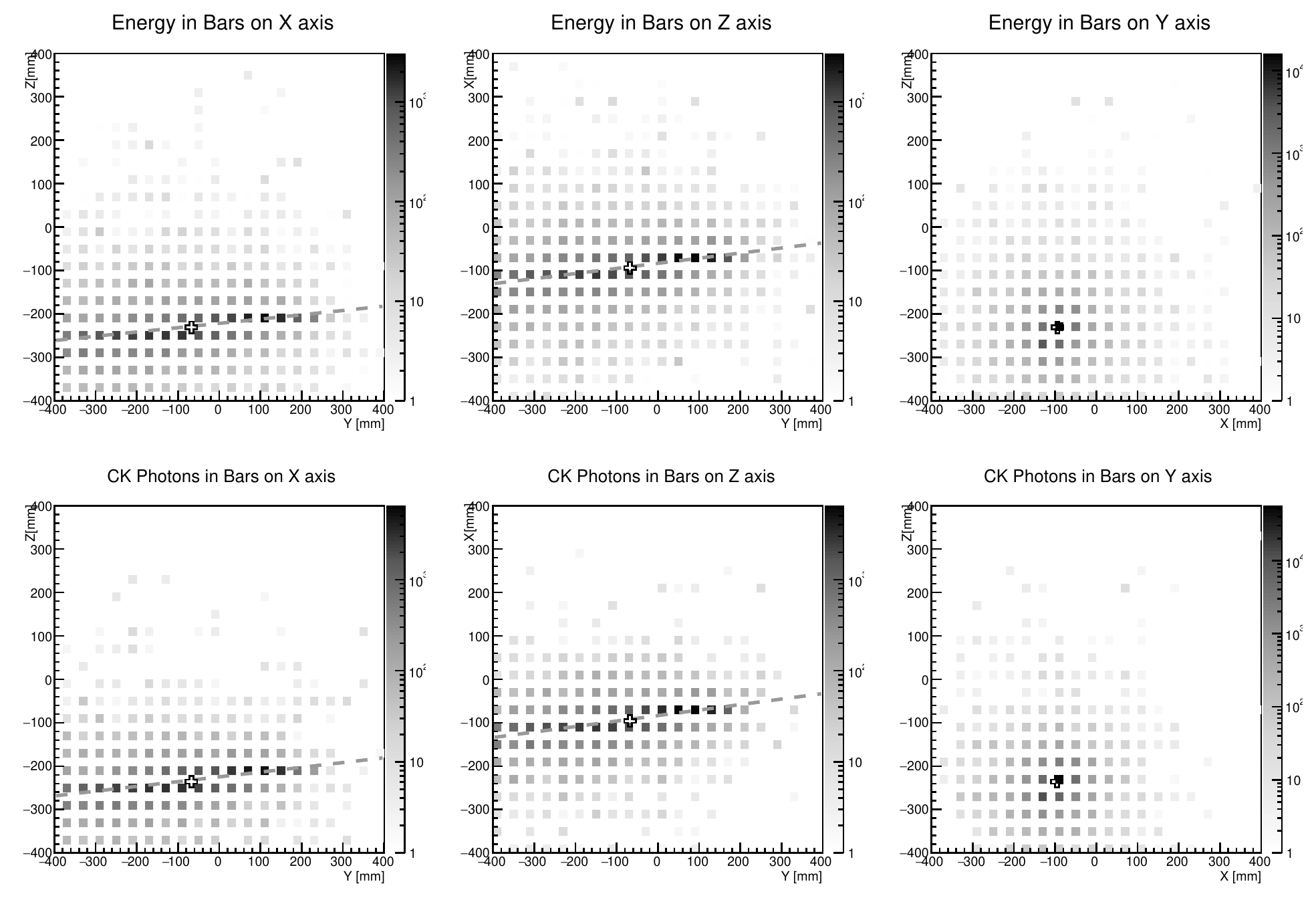}
\end{center}
\caption{Example of how a high-energy event looks like in this simulation:
distribution of S and C signals, for a single 2 TeV electron 
(top 2 rows) and 5 TeV proton (bottom 2 rows). The baricenter of the signals 
is shown with a cross, while the fits in the 2 lateral sides are shown with
gray dashed line.}\label{fig:evtdisplay}
\end{figure*}

We simulated a beam of electron and protons with 1/E spectrum 
from 20~GeV to 5~TeV. The particles enters from one side of the detector
with small incidence angle (up to 10$^{\circ}$). 
We used version 10.0-patch-04 of the Geant4 toolkit with the standard
FTFP\_BERT physics and optical photon generation turned on. 
Since the propagation of optical photon is very computing intensive, we killed 
Cherenkov photons tracks after generation and check their angle 
with respect to fiber direction to evaluate if they undergo internal reflection 
(in this case we consider the photon as collected in the sensor) or not. 
This solution permits to run a reasonable number of events with the available 
computing resources.

The reconstruction of particle direction is done in a very simple way, plotting 
the distribution of the S and C signal in the two lateral view 
(with respect to the beam direction), and fitting with a straight line
the two histograms. The fit slopes in the two directions are converted to 
coordinates in the detector reference frame and compared to the incoming 
particle direction. In detail, the beam enters from the $+Y$ surface, 
therefore the $XY$ and $ZY$ projections (which correspond respectively 
to bars along the $Z$ and along the $X$ direction) represent 
the two lateral views. The fit is done separately for S and C signal, 
so for each event, we have two direction estimates. 
Figure~\ref{fig:evtdisplay} shows an example
of this reconstruction for one event.

\begin{figure}[hbt]
\includegraphics[width=\linewidth]{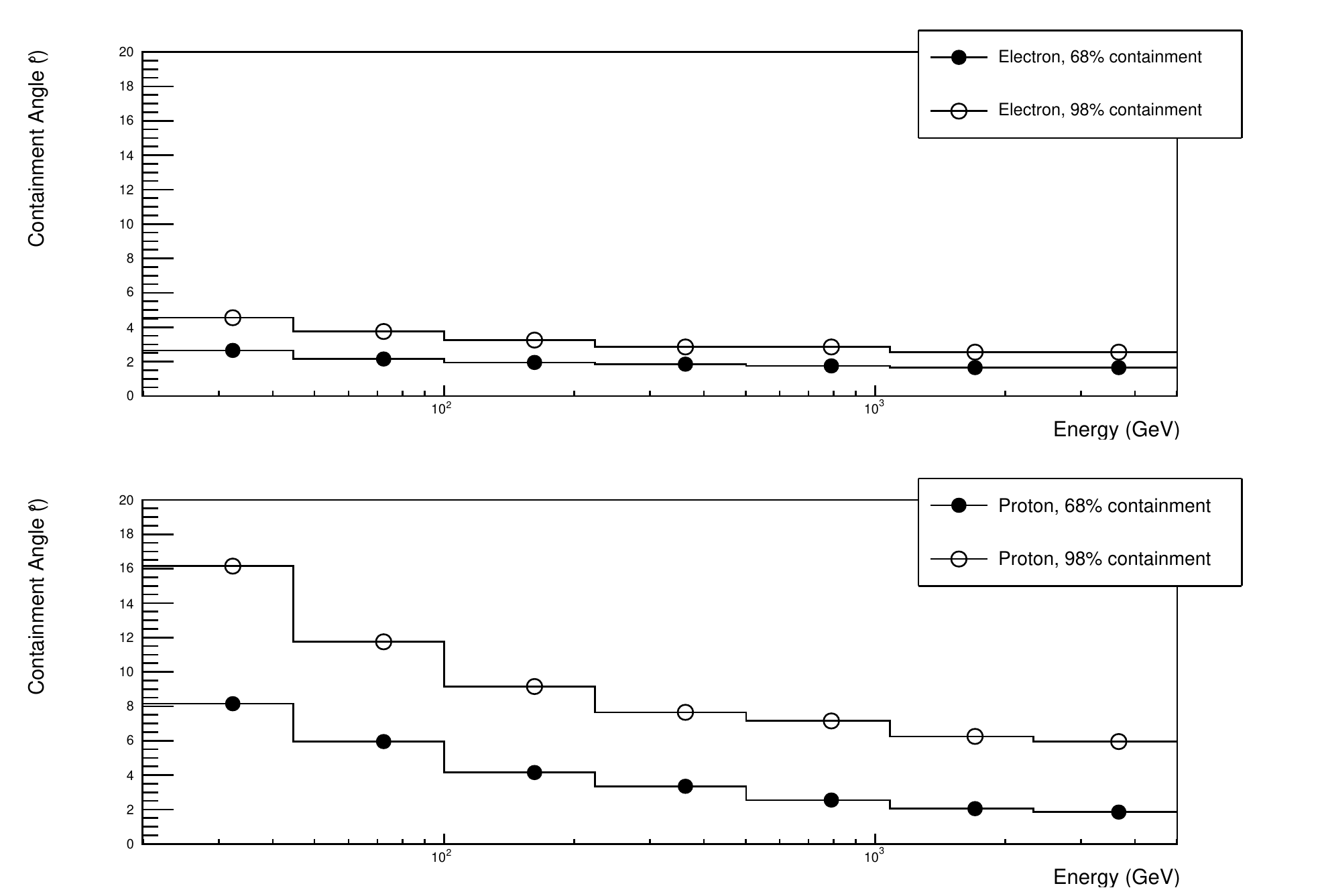}
\caption{Angular resolution (defined as  68\% and 95\% containment 
of true--reconstructed angle distribution) as function of energy, 
for electrons and protons.}\label{fig:psfvsenergy}
\end{figure}

To evaluate performance on particle direction we calculated the angle between 
true and reconstructed direction (Point Spread Function, PSF). 
The angular resolution is evaluated as the angle 
that contains 68\% or 95\% of the events. 
The results are shown in figure~\ref{fig:psfvsenergy} for S signal only, 
since the C signal was found to be almost equivalent and provides 
the same angular resolution, for both electron and proton and in the 
whole energy range. 
For electrons we see that the 95\% containment is quite close to the 68\% one,
indicating a compact distribution with small tail. The energy dependence is 
also small, the typical resolution is $\sim 2^{\circ}$.
For protons the tail is larger, increasing the value of the 95\% containment,
and there is also a larger energy dependence with the PSF improving with energy.
Even if it can be considered adequate for most of the high-energy cosmic-ray
science topic, it can be improved with a more clever reconstruction and
optimized design. As an example, each bar can be further subdivided in sectors
and read out with different devices to improve the granularity.

Particle energy is evaluated differently for electrons and protons, since
the two particle behaves very differently in the detector. 
Electrons have a very compact shower that is fully absorbed
in this detector.
Therefore the instrument works as total absorption calorimeter in this range
and the energy can be easily reconstructed by the sum of the S signal 
multiplied with a single calibration constant. 
The resulting energy resolution can easily be as low as few percent, 
and we won't discuss it further.

\begin{figure}[hbt]
%\centering
\includegraphics[width=\linewidth]{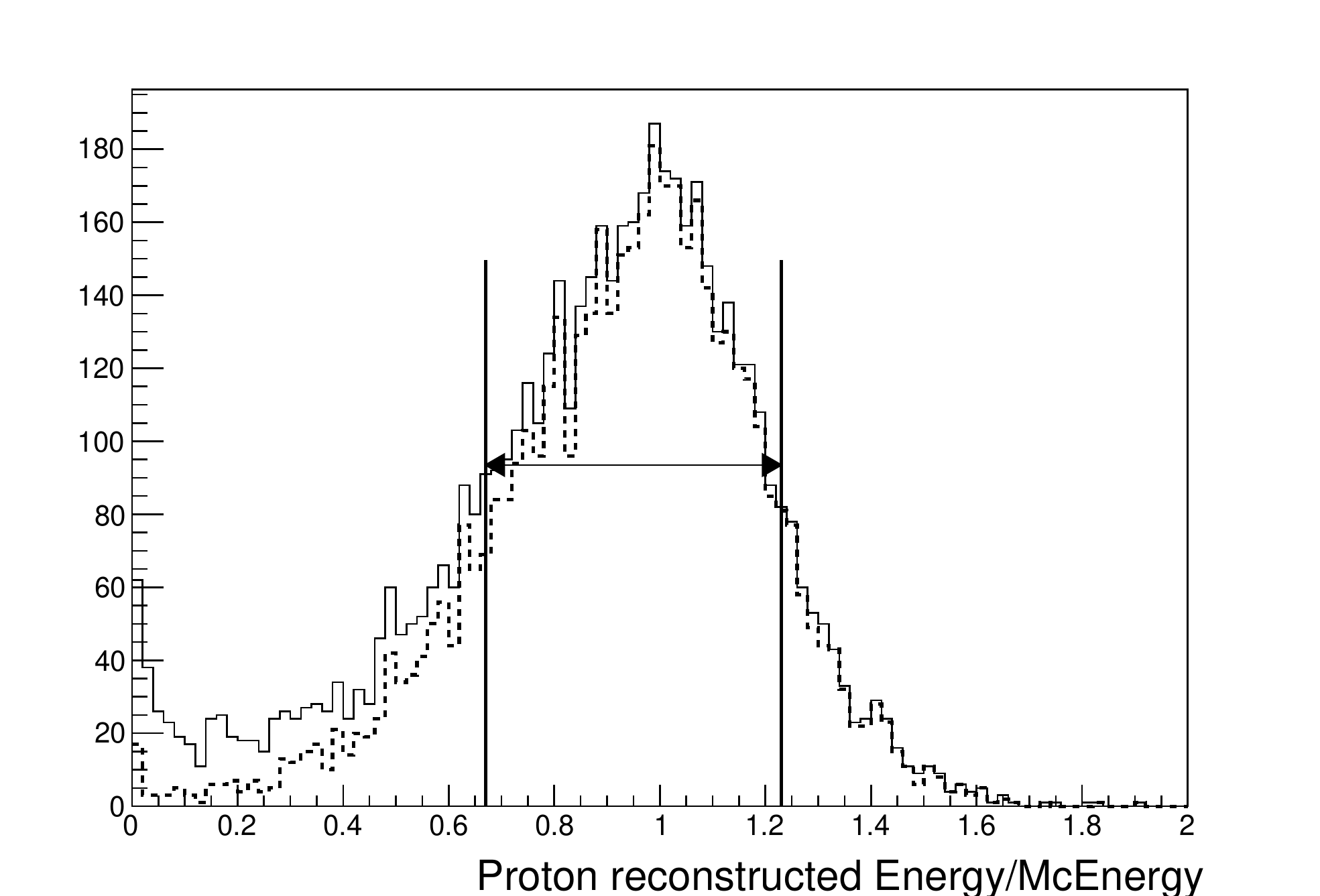}
\caption{Energy dispersion (reconstructed energy/true energy) for protons 
above 1 TeV. The arrow and the vertical lines show the windows that contains
68\% of the events, half of its width (the energy resolution) is about 28\%.
The dashed-line histogram represent the same distribution with a cut on C/S
ratio for a slightly better resolution of about 24\%.
}\label{fig:EneResPro}
\end{figure}

Proton energy reconstruction is more complex: hadronic shower is quite broad 
and starts later than the electromagnetic one. The S and C signal baricenter
distribution spans about the second half of the calorimeter depth,
and has small, but clear energy dependence.
As a consequence about half of the proton energy is lost outside the detector.  
The simple reconstruction algorithm implemented here uses the sum of S and C
signal with a single calibration constant each,
without trying to use shower shape information.
A small correction on the shower depth (the energy baricenter) is applied to 
the S sum, and the two signals are averaged together. 
The only event selection is to remove MIPs by requiring a minimum energy 
in the S signal empirically set to 100 MeV.
We studied the energy dispersion of this algorithm, defined as the
reconstructed energy divided by the true energy. 
As expected it peaks at around 1 and has a larger tail on the left side. 
Figure~\ref{fig:EneResPro} shows an example with all the events above 1~TeV.
The energy resolution is defined as the half-width of smallest window that
contains 68\% of the events (corresponding to $1\sigma$ in the gaussian case).
We found that we can achieve a value around 30--25\% in this energy range 
(with small energy dependency). A value that is close to the one
expected in the next generation instruments and that can be improved
with a more complex algorithm and detector optimization,
e.g. by a more careful selection of the passive material in terms of density,
radiation length and interaction length.

In this work we exploited the dual readout only for proton energy 
reconstruction, but there are several other advantages of this technique.
As an example, it has been shown that the ratio between C and S signal 
carry information on the electromagnetic fraction of the 
shower~\cite{dualreadout}. It can help to study the quality of 
the energy reconstruction and select events with good resolution in case 
it is important for specific science topics.
We can slightly improve the resolution with a cut
on a the C/S signal ratio reducing the left-side tail of the distribution
(dashed line in figure~\ref{fig:EneResPro}).
It can also be used for electron/hadron separation, together with other
shower topology information like its transverse size.
All these capabilities require a more advanced event reconstruction,
in particular in a complex geometry as the one 
we are proposing, for this reason we will not discuss them in details.

\section{Conclusions}
In this work we made a first description of a new detector for cosmic-ray 
study in orbit. The main idea is to have a modular detector that can scale up 
to a large size by keeping a good granularity and a relatively small number 
of channels. The implementation is done with detector elements that can be 
oriented along the 3 Cartesian axes and intersected in a way that provides
thee-dimensional sampling to particle showers. 

We tested this idea with simple simulations to make sure the concept works 
and to evaluate its performance in a realistic, although simplified, setup.
We found that we can achieve an acceptance of a few m$^2$ sr with a lateral 
size of the order of 1 m, even including the requirement of Earth-limited 
field of view. This is important since heavy satellites required by
this kind of science will likely be placed in Low Earth Orbits, 
where this condition holds. 
We showed that angular resolution of a few degrees is feasible and there is
room for improvements. We found that it can works not only as electrons, 
but also as protons telescope with good performance, with energy resolution 
down to $\sim 25\%$. 

Of course this work has to be considered as preliminary with some limitations
that we also discussed: the lack of sensor description in the simulation, 
the very simple event reconstruction etc. 
Other steps are necessary to move from an ideal concept to a feasible detector,
the most important would be to improve the simulation by implementing 
the missing parts and validate the concept with a detector prototype 
to be tested with real beam lines.

%%%%%%%%%%%%%%%%%%%%%%%%%%%%%%%%%%%%%%%%%%%%%%%%%%%%%%%%%%%%%
%%%%% References %%%%%

\bibliography{paper} 
%\bibliography{paper_biblio}   %>>>> bibliography data in report.bib
%\bibliographystyle{spiebib}   %>>>> makes bibtex use spiebib.bst

\end{document}